# Continual Prediction from EHR Data for Inpatient Acute Kidney Injury

Rohit J. Kate[a], Noah Pearce[b], Debesh Mazumdar[c], Vani Nilakantan[d]

[a]University of Wisconsin-Milwaukee, Milwaukee, WI, USA
[b]Aurora Research Institute, Aurora Health Care, Milwaukee, WI, USA
[c]Milwaukee Kidney Associates, Milwaukee, WI, USA
[d]Allina Health System, Minneapolis, MN, USA

**Abstract**

*Background*: Acute kidney injury (AKI) commonly occurs in hospitalized patients and can lead to serious medical complications. But it is preventable and if diagnosed and managed in time it is also potentially reversible. Hence several machine learning based predictive models have been built to predict AKI in advance from electronic health records (EHR) data. These models to predict inpatient AKI were always built to make predictions at a particular time, say 24 or 48 hours from admission; however, hospital stays can be several days long and AKI can develop any time within a few hours.

*Materials and Methods:* In order to optimally predict AKI before it develops at any time during a hospital stay, we present a novel framework in which AKI is continually predicted automatically from EHR data over the entire hospital stay instead of at only one particular time. The continual model predicts AKI every time a patient's AKI-relevant variable changes in the EHR. Thus the model is not only independent of a particular time for making predictions, but it can also leverage the latest values of all the AKI-relevant patient variables for making predictions. Using data of 44,691 hospital stays of duration longer than 24 hours we evaluated our continual prediction model and compared it with the traditional one-time prediction models.

*Results:* Excluding hospitals stays in which AKI occurred within 24 hours from admission, the one-time prediction model predicting at 24 hours from admission obtained area under ROC curve (AUC) of 0.653 while the continual prediction model obtained AUC of 0.724. The one-time prediction model that predicts at 24 hours obviously cannot predict AKI incidences that occur within 24 hours of admission which when included in the evaluation reduced its AUC to 0.57. In comparison, the continual prediction model had AUC of 0.709. The continual prediction model also did better than all other one-time prediction models predicting at other fixed times.

*Conclusion:* By being able to take into account the latest values of AKI-relevant patient variables and by not being limited to a particular time of prediction, the continual prediction model out-performed one-time prediction models in predicting AKI.

**Keywords**
acute kidney injury; prediction; EHR; machine learning

## 1. Introduction

Acute kidney injury (AKI), formerly known as acute renal failure, is a sudden loss of kidney function which affects 5-7% of hospitalized patients [1,2,3] and 22-57% of patients in intensive care unit [4,5,6,7,8]. AKI can lead to serious medical complications and is potentially fatal. It also results in longer hospital stays thus also increasing the associated healthcare cost [1]. Even after resolution, it can subsequently lead to severe kidney problems such as chronic kidney disease and progression to dialysis dependency. Incidence of AKI is highest in elderly patients [9,10] and its rate has been steadily increasing in this population due to increasing number of comorbidities, aggressive medical treatments and greater use of nephrotoxic drugs. AKI has a heterogeneous etiology and it often develops stealthily in hospitalized patients being treated for other problems; these two factors complicate its diagnosis. However, up to 30% of hospital-acquired AKI is preventable [11] if predicted in time. AKI is also potentially reversible if diagnosed and managed in time.

Its seriousness as a disease and its preventability together make AKI a perfect candidate for predictive analytics. Hence several machine learning based predictive models have been built to predict inpatient AKI from electronic health records (EHR) data [12,13]. All these models built in past to predict AKI to occur during rest of the hospital stay always had a particular time when the predictions were made, for example, at 24 hours after admission [14,15], or at 48 hours after admission [16,17], or around the time of a medical intervention [18,19,20]. However, there are two fundamental problems with such models that have a fixed time for predicting AKI. First, hospital stays of patients can be several days long during which their medical condition can significantly change and after which AKI may develop anytime within a few hours. In our data, for example, 39.4% of AKI incidences occurred after five days and 15.7% occurred after 10 days from admission. It is thus difficult to predict such incidences too far ahead in time. Second, if prediction model has a fixed time of prediction, say 24 hours after admission, then it is bound to miss all the AKI incidences that occur within 24 hours from admission. In our data, for example, 12.8% of AKI incidences occurred within 24 hours and 30.9% of AKI incidences occurred within 48 hours from admission. Thus the later the time of prediction of a model is the more incidences it will naturally miss. Hence neither a too early nor a too late prediction time is desirable for predicting AKI, instead, AKI should be predicted continually during the entire inpatient stay in order to optimally predict it before it develops. Although this possibility was recently mentioned in a workgroup statement from the 15$^{th}$ Acute Dialysis Quality Initiative (ADQI) consensus conference [12] as a model that would generate a prediction score in real time as each new data value is received, to the best of our knowledge no previous work ever built a model to continually predict AKI.

In this study, we introduce a novel framework for predicting AKI continually over hospital stays. A trained machine learning model is used to predict AKI every time a patient's status changes in the EHR, for example, when a new medication is prescribed, or a new comorbidity is recorded, or a new laboratory test result becomes available. We experimentally show that our continual prediction model obtains a superior prediction performance compared to the traditional one-time prediction models. In a recent study [21], an AKI prediction model was applied over different time zones of hospital stays in order to determine how early AKI can be predicted, for example, 24 hours or 48 hours before it occurs. However, their framework can only be used for retrospective evaluation and cannot be used in practice to predict AKI, say, 24 hours in advance, because one cannot know AKI's time of occurrence beforehand in order to apply their model 24 hours ahead of that time. In contrast, our continual prediction framework does not suffer from this problem and can be used in practice for predicting AKI in advance. It also does not require constant monitoring and is designed to work automatically using EHR data and trigger an alarm when necessary. Although we focused on AKI in this study, our proposed continual prediction framework is general and could also be applied to continually predict other diseases and disorders from EHR data.

We point out that our continual prediction framework is different from other well-known time-related data modeling frameworks. There are types of models that utilize repeated measures of variables from longitudinal data [22], however, they still make predictions at only one time. In contrast, our continual prediction model uses only the latest values of the variables but makes predictions continually over time. Survival analysis models [23] predict the time when an event will occur (e.g. time of death) but our continual prediction model continually predicts whether an event (e.g. AKI) will occur or not as the variables change over time. Multilevel models for change [24] model how variables change over time (e.g. to model growth), in contrast, our model is applied continually over time to predict a particular event. In past, researchers have done time series analysis of medical data for various applications [25,26,27,28,29] using the aforementioned types of models but they did not do continual prediction. Researchers have also built sequential models to discover longitudinal patterns in patient data, for example, patterns of disease diagnoses in order to predict the diagnosis for the next hospital admission [30,31]. In contrast, our model is not a sequential model but is applied continually over an entire hospital stay in order to predict a particular disease as early as possible using the latest values of patient variables.

## 2. Materials and Methods

### 2.1. Data Collection

The data was collected from the EHR of Aurora Health Care's 15 hospitals. All these hospitals are located in the southeastern region of Wisconsin state and use the same EHR system. Structured data along with their timestamps for entire hospital stays was collected for all adults older than 60 in the years 2013, 2014 and 2015 (number of patients=84,480). We focused on hospitalized older adults because occurrence of AKI is especially common in this age group [9,10]. Our exclusion criteria were patients who had chronic kidney disease stages III, IV and V, and those who were on dialysis. In addition, only hospital stays that were longer than 24 hours and in which at least two serum creatinine measurements were taken were included. After applying the exclusion criteria, there were 36,614 patients remaining with total 44,691 hospital stays among them as shown in Figure 1. This study was approved by the Institutional Review Board of Aurora Health Care.

### 2.2. AKI Definition

We used the AKIN criteria [32] as the definition of AKI. According to it, AKI is determined by either a 1.5 fold increase or an absolute increase of 0.3 mg/dL in serum creatinine (SCr) within 48 hours. In our data, AKI developed in 3,786 hospital stays. Please note that hospital stays are classified as AKI or non-AKI and not patients because a patient can have multiple hospital stays and may acquire AKI during one hospital stay and not during another. The time of the SCr measurement that satisfies the AKIN criteria is taken as the time of the AKI incidence. Figure 2 shows the number of AKI incidences that occurred by different times from admission.

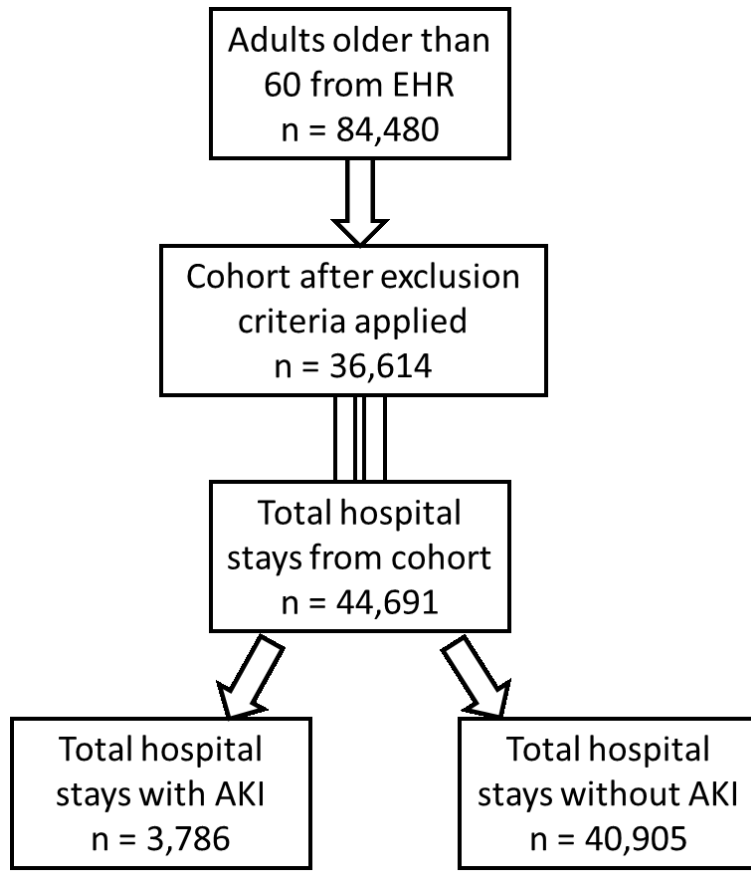

**Figure 1.** Flowchart depicting number of patients and hospital stays in the data.

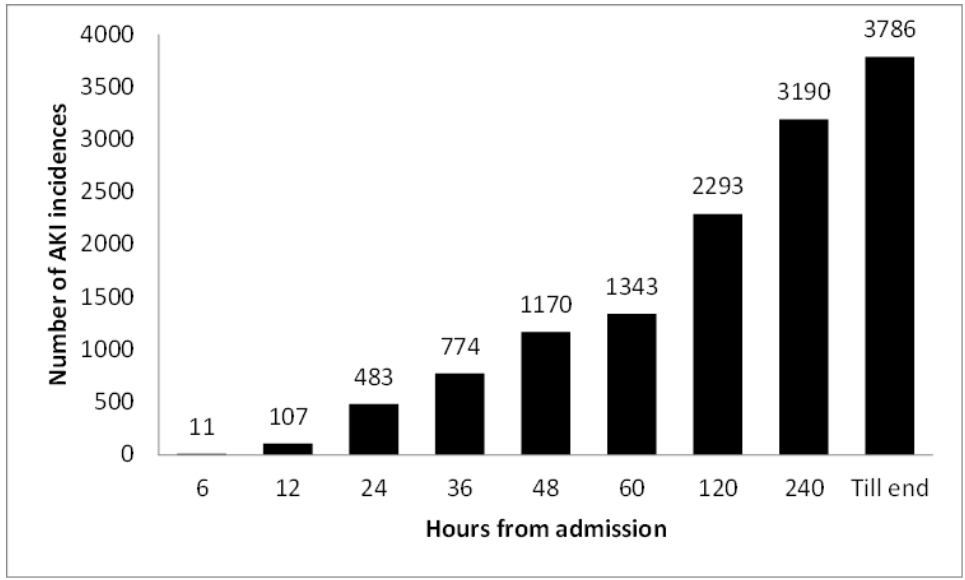

**Figure 2.** Number of AKI incidences by different hours from admission.

In the following subsections we first describe the traditional one-time prediction models for AKI and then we describe our new continual prediction model.

**2.3. Traditional One-Time Prediction Models for AKI**
The task of an AKI prediction model is to predict in advance whether AKI will develop or not during a patient's hospital stay. Machine learning methods have been used in past to build AKI prediction models that classify hospital stays as AKI or non-AKI from various predictive variables, also called features. A machine learning method is given training examples of hospital stays in the form of their features (described later) and the correct class (AKI or non-AKI). From this training data it learns to predict AKI on new hospital stays using only their features.

*2.3.1. Time of Prediction*
Most of the past work had chosen a particular time for making predictions, say 24 hours from admission [14,15], or 48 hours from admission [16,17], or even at the time of admission [33], while some of the past work were not clear about it [34,35]. In the rest of the paper, we call a prediction model built to make predictions at 24 hours from admission as *one-time-at-24-hour* prediction model. When such a model is trained, it uses values of the features as they were at 24 hours from admission for all its training examples. Then when such a model is applied to predict at 24 hours from admission whether AKI will develop during rest of the hospital stay, it uses values of the features as they were at 24 hours from admission during that hospital stay.

*2.3.2. Features*
Demographic information, comorbidities, medications and laboratory values are the common types of features that have been used to predict AKI. In this work, for both one-time and continual models, we used the same features that we had also used in our previous study [14]. These features had been decided with the help of a nephrologist who is a co-author. The only differences in features from our previous study is that in this study for each comorbidity and medication we used separate pre-admission and post-admission features and also used prior AKI as a feature. Table 1 shows the distribution of the feature values in our data across AKI and non-AKI hospital stays for the features that do not change over hospital stays. These features include all the demographic features and prior AKI.

**Table 1.** Distribution of feature values across AKI and non-AKI hospital stays in the data for the features that do not change over hospital stays. The p-values for statistical significance were computed using "N-1" chi-squared test [36] for proportions and using *t*-test for means.

| Variable | AKI (3786) | Non-AKI (40905) | *P*-value |
|---|---|---|---|
| Age | 72.99 ± 9.07 | 73.02 ± 9.25 | 0.85 |
| BMI | 29.42 ± 8.31 | 28.50 ± 7.32 | **< 0.01** |
| Gender = Male | 2031 (53.7%) | 20563 (50.3%) | **< 0.01** |
| Gender = Female | 1755 (46.3%) | 20342 (49.7%) | **< 0.01** |
| Race = White | 3340 (88.2%) | 37430 (91.5%) | **< 0.01** |
| Race = Black | 331 (8.8%) | 2440 (6.0%) | **< 0.01** |
| Race = Other | 115 (3.0%) | 1035 (2.5%) | 0.06 |
| Alcohol Use | 1014 (26.8%) | 13949 (34.1%) | **< 0.01** |
| Tobacco Use | 408 (10.8%) | 4771 (11.7%) | 0.09 |
| Drug Use | 69 (1.8%) | 472 (1.2%) | **< 0.01** |
| Prior AKI | 618 (16.3%) | 2838 (6.9%) | **< 0.01** |

Table 2 shows how the values of laboratory value, comorbidity and medication features are distributed across AKI and non-AKI hospital stays in our data. These features can change over hospital stays. Comorbidities and medications were binary features (true or false). Laboratory value features were the latest numeric values of the test results obtained since five days before admission till during the stay. To show that these features change over the course of hospital stays, we have shown the same statistics up to 24 hours following admission and after 24 hours following admission till AKI incidence or till the end for non-AKI hospital stays. For numeric features averages are shown. For this table, hospital stays in which AKI developed within 24 hours from admission (483) were excluded because they are not useful for comparing feature values up to 24 hours and after 24 hours following admission. Only post-admission comorbidity and medication features are shown because they can change over hospital stays. Please note that we did not use SCr as a feature because it is used as a gold-standard to determine AKI. In other words, the predictive model needs to predict AKI in absence of SCr measurement before the caregivers suspect the onset of AKI and order the SCr test. From the table it can be observed that for most features the differences between their values for AKI and non-AKI cases become more pronounced later during the hospital stays. For example, the difference in the use of diuretics between AKI and non-AKI cases was 13.74% up to 24 hours following admission while it was 30.12% after 24 hours following admission. This shows that the status of patients change significantly over hospital stays and a one-time prediction model (e.g. at 24 hours) will not be able to take this into account for predicting AKI.

### 2.3.3. Evaluation

The standard way to evaluate a machine learning model is through ten-fold cross-validation [37]. In this process the data is first randomly divided into ten equal parts. Then in each fold, nine parts are used for training the model and the remaining part (different in each fold) is used for testing it. The results of all the folds are combined and reported. Sensitivity (proportion of positive examples correctly identified) and specificity (proportion of negative examples correctly identified) are the most common evaluation measures for predictive models. Many machine learning methods give probability (or confidence) of their prediction and by varying the threshold value (we used step size of 0.01) on a model's prediction confidence, one can trade-off sensitivity with specificity and thus generate an entire curve between them. Traditionally, the curve is plotted between sensitivity (true positive rate) and 1-specificity (false positive rate) and is called an ROC curve. The area under this curve (AUC) is one number that conveniently summarizes the entire curve. In our results, we report AUC and use it for comparing different models.

**Table 2.** Distribution of feature values across AKI and non-AKI hospital stays up to 24 hours and after 24 hours following admission for the features that can change over hospital stays. The hospital stays in which AKI developed within 24 hours of admission (483) were excluded for the purpose of this table. The differences shown in bold were found to be statistically significant (p < 0.05) using "N-1" chi-squared test [36] for proportions and using *t*-test for means.

|  | Up to 24 hours following admission | | | After 24 hours following admission | | |
|---|---|---|---|---|---|---|
|  | AKI (3,303) | Non-AKI (40,905) | Difference | AKI (3,303) | Non-AKI (40,905) | Difference |
| *Laboratory Values (Means, Standard deviations)* | | | | | | |
| AST | 29.99 ±18.12 | 26.41 ± 15.25 | **3.58** | 31.27 ± 17.85 | 29.05 ± 17.79 | **2.22** |
| Blood Bilirubin | 0.6 ± 0.37 | 0.57 ± 0.34 | **0.03** | 0.6 ± 0.37 | 0.58 ± 0.36 | **0.02** |
| BP Diastolic | 68.65 ± 10.5 | 70.25 ± 9.64 | **-1.6** | 65.64 ± 8.38 | 68.68 ± 8.31 | **-3.04** |
| BP Systolic | 131.22 ±19.76 | 133.89 ± 18.43 | **-2.67** | 125.91 ± 15.79 | 130.48 ± 15.81 | **-4.57** |
| BUN | 20.55 ± 9.67 | 18.58 ± 7.66 | **1.97** | 21.64 ± 9.5 | 17.86 ± 7.73 | **3.78** |
| Heart Rate | 82.31 ±14.61 | 79.8 ± 14.39 | **2.51** | 81.65 ± 12.79 | 77.79 ± 12.17 | **3.86** |
| Platelet Count | 220.52 ±87.79 | 219.87 ± 79.09 | 0.65 | 217.41 ± 88.04 | 212.36 ± 80.65 | **5.05** |
| Temperature | 98.16 ± 0.8 | 98.14 ± 0.64 | 0.02 | 98.22 ± 0.62 | 98.13 ± 0.48 | **0.09** |
| Troponin | 0.33 ± 0.72 | 0.29 ± 0.65 | 0.04 | 0.38 ± 0.82 | 0.38 ± 0.8 | 0 |
| *Comorbidities (Percentages)* | | | | | | |
| Coronary Artery Disease | 0 | 0 | 0 | 0 | 0 | 0 |
| Diabetes | 5.21 | 4.36 | **0.85** | 6.96 | 5 | **1.96** |
| Disorders of Lipoid Metabolism | 1 | 1.68 | **-0.68** | 1.94 | 2.26 | **-0.32** |
| Heart Failure | 12.72 | 5.36 | **7.36** | 18.07 | 6.98 | **11.09** |
| Hypercalcemia | 0.36 | 0.19 | 0.17 | 0.67 | 0.26 | **0.41** |
| Hyperlipidemia | 2.27 | 3.6 | **-1.33** | 2.76 | 4.09 | **-1.33** |
| Hypertension | 26.73 | 30.4 | **-3.67** | 40.33 | 37.79 | **2.54** |
| Pancreatitis | 0.97 | 1.22 | -0.25 | 1.27 | 1.35 | -0.08 |
| Respiratory Failure | 1.94 | 1.05 | **0.89** | 3.39 | 1.35 | **2.04** |
| Rhabdomyolysis | 0.42 | 0.34 | 0.08 | 0.45 | 0.38 | 0.07 |
| Sepsis | 5.33 | 3.69 | **1.64** | 7.36 | 4.31 | **3.05** |
| Thrombocytopenia | 1.88 | 1.14 | **0.74** | 3.48 | 1.84 | **1.64** |
| *Medications (Percentages)* | | | | | | |
| ACE Inhibitors | 17.62 | 16.06 | **1.56** | 26.1 | 19.05 | **7.05** |
| ACE Inhibitors or NSAIDS or Diuretics | 46.26 | 38.59 | **7.67** | 71.84 | 47.51 | **24.33** |
| ACE Inhibitors or ARB or NSAIDS or Diuretics | 49.35 | 43.38 | **5.97** | 74.72 | 51.88 | **22.84** |
| Acylovir | 1.18 | 0.79 | **0.39** | 2.36 | 1.11 | **1.25** |
| Aminoglycosides | 0.27 | 0.08 | **0.19** | 0.54 | 0.16 | **0.38** |
| ARB | 8.39 | 8.88 | -0.49 | 12.44 | 9.96 | **2.48** |
| Cisplatin | 0.33 | 0.02 | **0.31** | 0.7 | 0.07 | **0.63** |
| Diuretics | 35.15 | 21.41 | **13.74** | 59.88 | 29.76 | **30.12** |
| K Sparing | 5.24 | 2.37 | **2.87** | 8.08 | 3.28 | **4.8** |
| Lipid Lowering Drugs | 37.63 | 39.03 | -1.4 | 50.89 | 47.34 | **3.55** |
| NSAIDS | 5.24 | 9.14 | **-3.9** | 10.81 | 11.92 | **-1.11** |
| Radiocontrast Dyes | 16.65 | 18.15 | **-1.5** | 27.52 | 22.57 | **4.95** |

## 2.4. Continual Prediction Model for AKI

We first define three terms *event*, *event-time* and *feature-snapshot* that will be useful in describing our new continual prediction model. We call *event* as any change in the value of any of the features during a hospital stay. For example, an event occurs any time a new medication is prescribed, or a new comorbidity is recorded, or a new laboratory test result becomes available. If multiple of these changes occur at the same time (to be more precise, if multiple of these get recorded in the EHR with the same timestamp) then all are considered as one event. Admission and discharge are also regarded as events. We call *event-time* as the timestamp of an event as recorded in the EHR. Finally, we call the latest values of all the features at an event-time as its *feature-snapshot*. Figure 3 (a) shows a hypothetical illustrative timeline of a hospital stay with some events and their corresponding event-times and feature-snapshots. In our data, there were on average 21.5 (standard deviation=12.6) events per day per hospital stay.

### 2.4.1. Applying the Model

Once trained (we later describe how it is trained), our continual prediction model is applied to predict AKI at every event-time of a hospital stay starting from the admission time. There is no need to apply the model between two consecutive events because by definition no feature changes between two consecutive events therefore the model's prediction will not change. To apply the model at an event-time, the values of the features are taken from that event-time's feature-snapshot.

Many machine learning methods give probability of their prediction, for example, in our task it is the probability that AKI will develop. Thus the continual prediction model's predicted probability that AKI will develop dynamically changes during the course of a hospital stay as the feature values change. Figure 3 (b) illustrates this for the example hospital stay shown above it. To contrast this, Figure 3 (c) shows the static nature of the one-time-at-24-hour prediction model's prediction. Unlike one-time prediction models, the continual prediction model is able to take into account the latest medical status of the patient as it changes dynamically over the hospital stay. This helps the model in predicting AKI incidences more accurately that develop later in the hospital stay. Additionally, given that the continual prediction model is applied starting from the admission time, it cannot miss early AKI incidents the way a one-time prediction model misses those because the model is applied too late. Thus the continual prediction model circumvents the two problems of one-time prediction models pointed out earlier of either being applied too early to accurately predict AKI or being applied too late to prevent AKI.

Although the continual prediction model gives probability of AKI at every event-time, it will be highly impractical to alert healthcare providers about the probability of AKI every time a feature changes. Hence an important question is – when is it appropriate to trigger an alarm for AKI as predicted by the model? In our method, we use a threshold (a parameter) and whenever the AKI prediction probability exceeds this threshold it is deemed that the model has predicted AKI which should be alerted to the providers. Figure 3 (b) shows an example of AKI being predicted by the continual prediction model using a particular threshold value of 0.8 (this threshold value was arbitrarily chosen for illustration; for evaluation, threshold values between 0 and 1 in step-size of 0.01 were used to obtain different points on the ROC curve). The first event-time at which this happens is recorded as the time the model predicted AKI for that hospital stay. If the probability never exceeds the threshold over a hospital stay then it is deemed that the model did not predict AKI for that hospital stay.

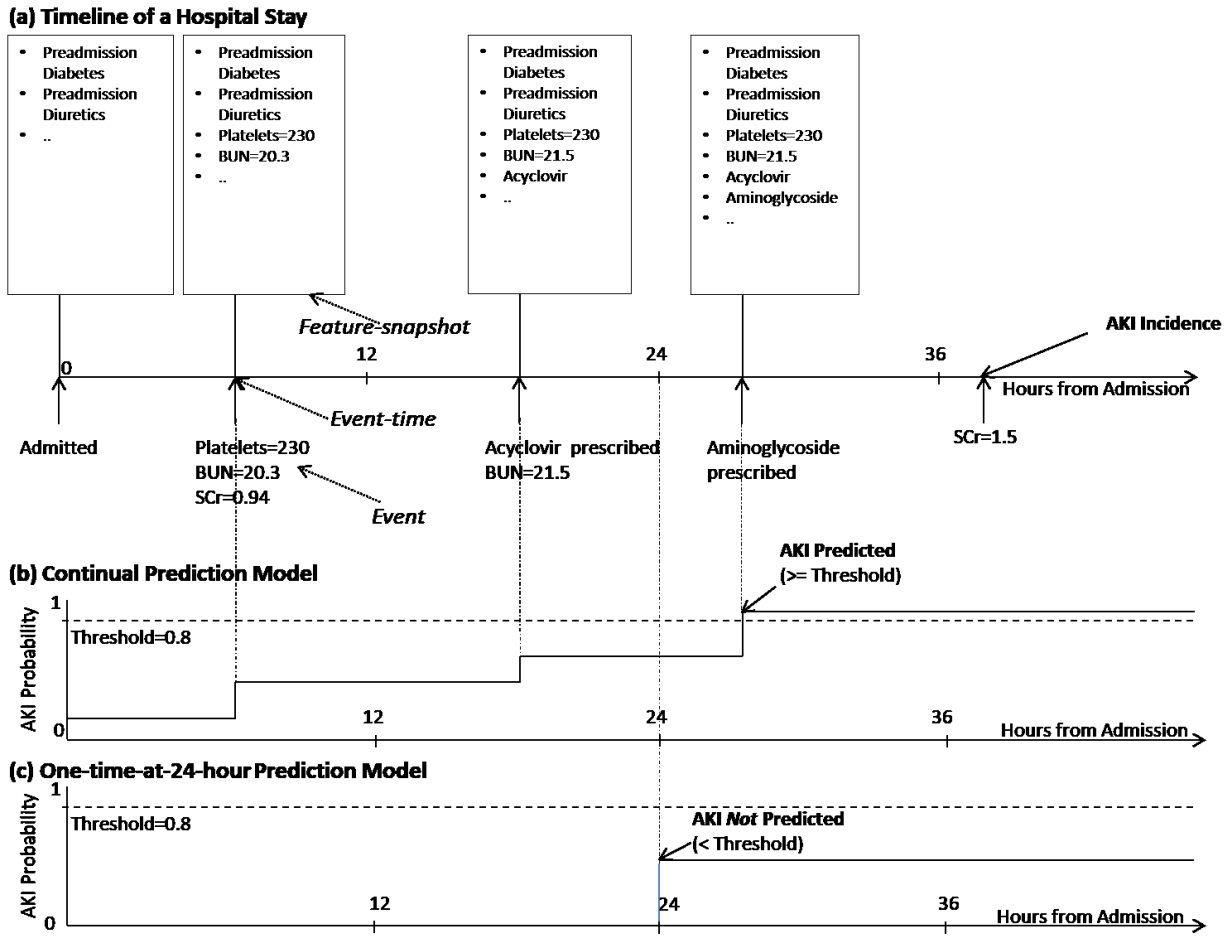

**Figure 3.** An illustrative timeline of a hospital stay along with the AKI prediction probabilities of the continual and the one-time-at-24-hour prediction models. The continual prediction model predicts AKI once aminoglycoside is prescribed sometime after 24 hours from admission but the one-time-at-24-hour prediction model fails to predict AKI at 24 hours from admission. AKI is determined sometime after 36 hours by the AKIN criteria. This is a hypothetical example for illustration purpose.

In the hypothetical illustrative example shown in Figure 3, AKI develops after aminoglycoside is prescribed sometime after 24 hours from admission. The continual prediction model is able to predict AKI when aminoglycoside is prescribed (the AKI probability increases and exceeds the threshold). On the other hand, the one-time-at-24-hour prediction model fails to predict AKI because aminoglycoside was prescribed after 24 hours from admission.

### 2.4.2. Evaluation

For a hospital stay during which AKI developed, the continual prediction model must predict AKI (based on the threshold) by the time of AKI incidence otherwise it will be considered that the model missed it (predicting it later will be too late). In other words, to correctly predict AKI for a hospital stay, the continual prediction model's prediction probability for AKI must be above the chosen threshold at least at one event-time before the time of AKI incidence. On the other hand, for a hospital stay during which AKI did not develop, the continual prediction model must not predict AKI at any event-time. In other words, to correctly predict non-AKI for a hospital stay, the continual prediction model's prediction probability for AKI should not be above the chosen threshold at *any* event-time over the entire hospital stay. Please note that this is a very strict criterion for correct non-AKI prediction and given that the number of non-AKI hospital stays are more than 10 times the number of AKI hospital stays in our data this makes the task for the continual prediction model especially difficult. The threshold offers a trade-off between sensitivity and specificity. A higher threshold will increase the model's specificity but decrease its sensitivity and vice-versa. The rest of the evaluation proceeds in the same way as for the one-time prediction models as described earlier.

*2.4.3. Training the Model*

For training the one-time-at-24-hour prediction model, it is clear that the feature values for positive (AKI) and negative (non-AKI) training examples must be collected at 24 hours from admission. However, for the continual prediction model, which has no specific time of prediction, an important question arises – what should be the time for collecting the feature values for positive and negative training examples? Ideally, for each hospital stay in the training data during which AKI developed, feature-snapshot of its each event-time before AKI should be used to create a positive example because when the model is applied it is expected to predict AKI during those events (sooner the better). Similarly, for each hospital stay in the training data during which AKI did not develop, feature-snapshot of its each event-time till the end of hospital stay should be used to create a negative example. However, this will result in an explosion of the number of training examples (0.48 million positive and 3.2 million negative examples in our data) which would require unreasonable amounts of computational time and memory for training machine learning methods. Also, any two consecutive feature-snapshots will minimally differ from each other hence most of the examples thus created will be largely repetitious.

In this work, we decided to take feature-snapshot at the time AKI was determined (i.e. when the SCr measurement satisfied the AKIN criteria) for creating positive training examples, and at the time last SCr measurement was taken for creating negative training examples. This way we create only one training example per hospital stay thus keeping under control the computational time and memory requirement for training machine learning methods. We decided to use the time of AKI for creating positive training examples because it best represents the status of patients who have acquired AKI. We decided to use the time of last SCr for creating negative training examples because it shows that the patient was still considered prone to AKI. This way the model will learn to better distinguish between AKI and non-AKI cases. If we had used, for example, feature-snapshot of the discharge time for creating negative examples then the model may simply learn to predict whether the patient is about to get discharged or not. We also tried random event-times to create negative examples but that did not lead to any noticeable difference in the results. We want to point out that training a prediction model with the examples as described above is similar to the "AKI detection" model we had built in our past work [14]. However, that model was not applied continually and its purpose was to determine if AKI had already developed in order to prevent it from going undiagnosed.

In our previous work [14] we had compared several machine learning methods for building AKI prediction models and had found logistic regression [38] to work best. Given that the goal of the current study was to compare the proposed continual prediction model with the traditional one-time prediction models, we used only logistic regression to build both types of models. Comparing different machine learning methods was not a goal of this study which we had done in our previous work. We used the Weka machine learning software [39] to build logistic regression models. Missing feature values were handled by Weka's default mechanism for logistic regression in which missing values are replaced by the modes for nominal features and by the means for numeric features as computed from the training data. Given the unbalanced nature of our data with more than 10 times non-AKI examples than AKI examples, we used Weka's cost-sensitive classifier whose weight for the minority class was decided out of 1, 2, 4, 6, …, 18, 20 through internal cross-validation within the training data.

## 3. Results and Discussion

### 3.1. Continual vs. One-time Prediction Models

Table 3 shows comparison between AUC obtained by the continual prediction model and by the one-time-at-24-hour prediction model through 10-fold cross-validation using exactly the same folds. The first row shows results in which we exclude hospital stays in which AKI developed within 24 hours from admission (483 such hospital stays) because it is beyond the capacity of the one-time-at-24-hour prediction model to predict these AKI incidences. Continual prediction model obtained statistically significantly better AUC than the one-time prediction model (0.724 vs. 0.653; $p < 0.05$; two-tailed paired *t*-test).

In the second row we show results for all hospital stays. In this comparison, all the 483 hospital stays in which AKI developed within 24 hours will be missed by the one-time-at-24-hour prediction model. As a result, its AUC dropped to 0.57 while the AUC of the continual prediction model was 0.709. Figure 4 shows the corresponding ROC curves. Overall, this is a more realistic comparison because it does not artificially exclude the cases which are beyond the capacity of the one-time prediction model. However, in past the results for one-time prediction models were always reported after excluding these cases, hence those results looked better than how they were in reality. It should be noted that the ROC curve for the one-time-at-24-hour prediction model in Figure 4 does not end at the top right corner. The reason for this is that even with zero threshold value for confidence, which usually yields maximum sensitivity of 1, the one-time-at-24-hour model simply cannot predict those 483 AKI hospital stays because the prediction time of 24 hours is already too late for them.

**Table 3.** Comparison of continual prediction model and one-time-at-24-hour predcition model for AKI. An AUC number in bold was found to be statistically significant compared to the AUC number in the same row ($p < 0.05$; two-tailed paired *t*-test).

|  | AKI Hospital Stays | Non-AKI Hospital Stays | Continual Prediction Model AUC (95% CI) | One-time-at-24-hour Prediction Model AUC (95% CI) |
|---|---|---|---|---|
| Excluding hospital stays in which AKI developed within 24 hours from admission | 3303 | 40905 | **0.724** (0.705, 741) | 0.653 (0.641, 665) |
| All hospital stays | 3786 | 40905 | **0.709** (0.690, 0.728) | 0.57 (0.555, 0.584) |

In order to show that the continual prediction model is also better than other one-time prediction models besides the one-time-at-24-hour prediction model, we generated results for different one-time prediction models that predict from 0 hours (i.e. at the time of admission) to 120 hours from admission. Each of these models requires separate training and testing for its time of prediction. The second column of Table 4 shows the results for all hospital stays. The next column shows results of evaluation after excluding hospital stays in which AKI developed before the time of prediction as well as after excluding hospital stays which were shorter than the time of prediction. Finally, for a direct comparison of all the models on the same set of hospital stays on which none of them was applied too late, results are also shown in the last column after excluding hospital stays in which AKI developed within 120 hours as well as after excluding hospital stays which were shorter than 120 hours. For comparison, the last row shows the corresponding results of the continual prediction model.

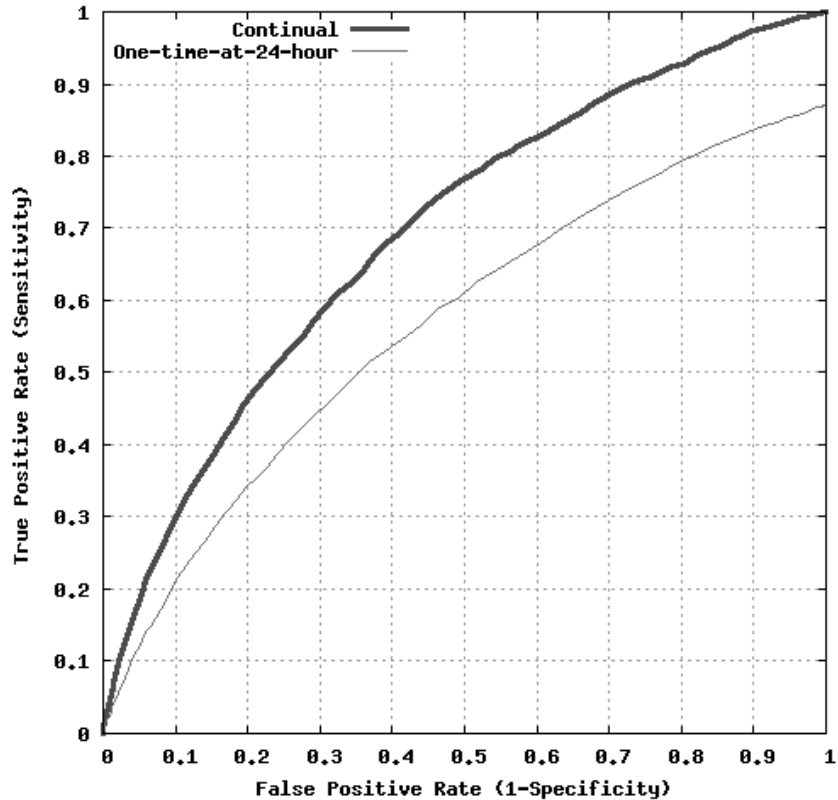

**Figure 4.** ROC curves of the continual prediction model and the one-time-at-24-hour prediction model when evaluated on all hospital stays.

It can be seen from Table 4 that none of the one-time prediction models does better than the continual prediction model. As we had pointed out earlier, if the one-time prediction is made too late then it would miss AKI cases that developed earlier. The drop in AUC in the second column clearly shows this. It is interesting to note that the last column of Table 3 changes very little over different times of one-time prediction models indicating that a later one-time prediction model is not necessarily better at predicting AKI incidences that develop later. This may seem counter-intuitive, but it has an explanation. Using information gain statistic [40] we show in the Appendix that the pre-admission features are the most important features for one-time prediction models. This is because these models are supposed to predict far ahead in time and hence the other features that later change over hospital stays are not helpful in making such early predictions. Hence all the one-time prediction models predominantly use pre-admission features to make predictions. And given that the pre-admission features do not change over hospital stays, a later one-time prediction model is not necessarily better at predicting AKI that develop later. We also show in the Appendix that, in contrast, for the continual prediction model the features that change over hospital stays are the most important features.

**Table 4.** Comparison of one-time prediction models with different times of predictions. All evaluation numbers are AUC.

| One-time prediction at X hours  X = | All hospital stays (AKI developed within X hours will be considered missed) (AKI: 3786; Non-AKI: 40905) | Excluding hospital stays which were shorter than X hours or in which AKI developed within X hours | Excluding hospital stays which were shorter than 120 hours or in which AKI developed within 120 hours (AKI: 1493; Non-AKI: 12274) |
|---|---|---|---|
| 0 | 0.653 | 0.653 | 0.597 |
| 12 | 0.637 | 0.655 | 0.594 |
| 24 | 0.570 | 0.653 | 0.592 |
| 36 | 0.560 | 0.650 | 0.593 |
| 48 | 0.525 | 0.641 | 0.595 |
| 60 | 0.543 | 0.629 | 0.595 |
| 72 | 0.535 | 0.634 | 0.607 |
| 84 | 0.559 | 0.628 | 0.607 |
| 96 | 0.553 | 0.624 | 0.609 |
| 108 | 0.566 | 0.611 | 0.607 |
| 120 | 0.560 | 0.607 | 0.607 |
|  |  |  |  |
| Continual | 0.709 | - | 0.651 |

### 3.2. Advance Prediction

One can also evaluate retrospectively how far early could a model predict AKI before it developed. This may be desirable to know in order to have sufficient time to respond to prevent AKI. In Figure 5, we show how AUC varied with the number of hours in advance a model must predict AKI before it developed, i.e. it may predict earlier but if it predicted any later then it will be considered too late and the prediction will not count as correct. For example, if the model must predict AKI at least 12 hours in advance then its time of prediction must be 12 hours before AKI developed. For the one-time-at-24-hour prediction model the time of prediction is fixed at 24 hours, but for the continual prediction model the time of prediction is the first event-time when the AKI probability exceeds the threshold (see Figure 1 illustration). For this plot, we only considered AKI incidences that developed at least 24 hours after admission for a fair comparison across all time lengths from 0 to 24 hours (for example, if AKI developed in less than 24 hours then one could not have possibly predicted it 24 hours in advance). We plotted the curves with 2 hour granularity which was sufficient to obtain smooth curves, they can be otherwise plotted at any granularity level. Please note that the continual prediction model still makes predictions continually, here we are only evaluating in 2 hour intervals how far in advance it could predict AKI (i.e. AUC for prediction by 6 hours in advance, AUC for prediction by 8 hours in advance etc.). As one would expect, the figure shows that for both types of models it is harder to predict AKI more time in advance. The AUC seems to slowly drop almost linearly with the number of hours in advance by when the prediction is required. However, it can be seen that the performance of the one-time-at-24-hour model drops more precipitously than the continual prediction model.

It can be seen from Figure 5 that the AUC of the one-time-at-24-hour prediction model for making prediction by 24 hours in advance is nearly 0.5. A random prediction model also gives AUC of 0.5 with a diagonal ROC curve. However, we want to point out that there is a different reason why the one-time-at-24-hour prediction model obtains such low AUC for advance prediction. When the one-time-at-24-hour prediction model needs to predict AKI by 24 hours in advance, it simply cannot predict AKI incidences that occur between 24 hours and 48 hours of admission. Hence those incidences become out of reach for the model to predict and that lowers the maximum sensitivity it can obtain. Consequently, its ROC curve does not reach the top right corner. A similar situation was shown in Figure 4 for a curve that had AUC of 0.57. It should be noted that in this situation a random prediction model applied to predict at 24 hours will obtain AUC lower than 0.5 (its ROC curve will not reach the top right corner either), hence the one-time-at-24-hour learned prediction model is still doing better than a random prediction model. Of note, because continual prediction model has no fixed time of prediction, there are no AKI incidences which are beyond its reach to predict in advance. For instance, when the continual prediction model needs to predict AKI by 24 hours in advance, it is possible for it to predict AKI incidences that occur between 24 hours and 48 hours of admission by making those predictions any time between 0 and 24 hours of admission.

We want to contrast our evaluation with a recent study [21] in which in order to evaluate the performance of an AKI prediction model to make predictions, say 24 hours in advance, the model was actually applied 24 hours before each AKI incidence. As was pointed out in the Introduction, their framework of advance prediction is not possible to deploy in practice because one cannot know in advance the time of an AKI incidence in order to determine when the model should be applied (in fact, that defeats the very purpose behind the prediction model). In contrast, in our evaluation, a model makes prediction when it makes prediction (as depicted in the illustrative example in Figure 1) and it is then evaluated whether that prediction was made, say, by 24 hours in advance of the AKI incidence. Hence our evaluation does not require knowing the time of an AKI incidence beforehand in order to determine when to predict it. Our framework is thus deployable in practice for advance prediction and our evaluation also realistically measures the advance prediction performance.

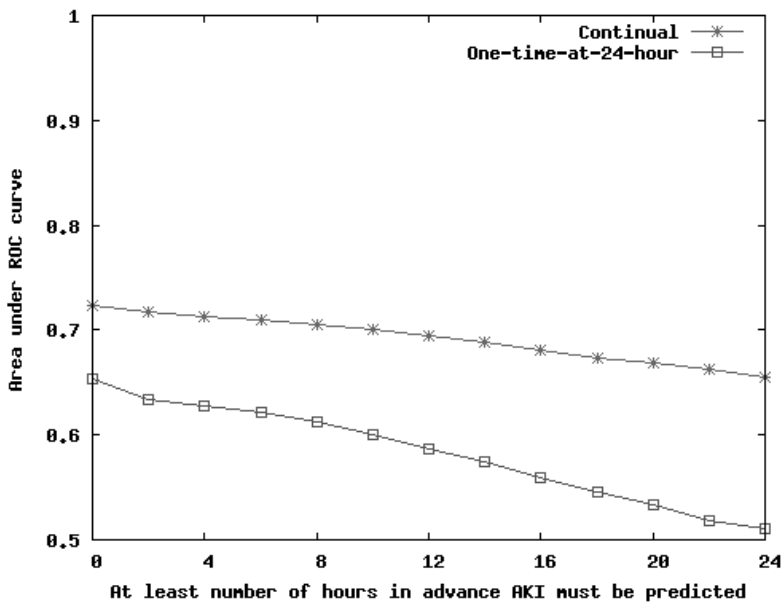

**Figure 5.** A graph showing how the prediction performance changed with the number of hours in advance by when a model must predict AKI.

### 3.3. Dynamic Predictive Features

Another advantage of our continual prediction framework is that we can directly observe how the model's prediction dynamically changes as the values of the features change over the course of hospital stays. The third column of Table 5 shows the percentage of hospital stays over which a particular medication was prescribed or a particular comorbidity was diagnosed. The fourth (last) column shows how often the model's prediction changed from non-AKI to AKI whenever the medication was prescribed or the comorbidity was diagnosed over the hospital stay (it is possible that other features could have also changed at the same time). One can compute these numbers for every confidence threshold but we are showing the combined results for all confidence thresholds instead of picking any particular threshold. The table shows only the top ten such features. Laboratory values are numeric features that can change multiple times during a hospital stay in different magnitudes hence they could not be included in this analysis. It can be seen from the table that prescription of medications cisplatin and aminoglycosides were most prominently associated with change in prediction from non-AKI to AKI. The third column shows that these were not prescribed frequently though. The next two medications to follow were diuretics and acyclovir. All these medications are, in fact, known for their nephrotoxicity. Hypercalcemia and respiratory failure were found to be the most prominent among comorbidities which when diagnosed during hospital stay were most often associated with change in prediction from non-AKI to AKI. Such dynamic information was never reported in the past work and could not have been obtained without a continual prediction model.

**Table 5.** Top ten medications and comorbidity features whose prescription or diagnosis during hospital stay were associated with change in the model's prediction from non-AKI to AKI. The numbers in the last column are average values obtained over all confidence thresholds.

| Rank | Medication or Comorbidity Feature | Percentage of hospital stays over which the feature changed (%) | How often change in feature over hospital stay changed prediction from non-AKI to AKI (%) |
|---|---|---|---|
| 1 | Cisplatin | 0.123 | 25.6 |
| 2 | Aminoglycosides | 0.186 | 15.4 |
| 3 | Hypercalcemia | 0.311 | 5.81 |
| 4 | Diuretics | 36.4 | 3.84 |
| 5 | Acyclovir | 1.27 | 3.6 |
| 6 | ACE inhibitors or NSAIDS or Diuretics | 56.4 | 2.53 |
| 7 | ARB or ACE inhibitors or NSAIDS or Diuretics | 61.5 | 2.29 |
| 8 | Respiratory failure | 1.59 | 2.17 |
| 9 | Rhabdomyolysis | 0.403 | 2 |
| 10 | K sparing | 4.19 | 1.86 |

### 3.4. Limitations and Future Work

Our data was limited to one system of hospitals which limits generalization. The data was obtained only from the structured part of EHR. In future, natural language processing techniques [41] could be used to extract AKI-relevant features and their values from the text part of EHR. Our continual prediction model always used the latest values of all the features; in future, it could be improved by also taking into account the past values of the features and how they vary over time. Our study was retrospective and it remains to be seen how such a framework will perform when deployed. However, we point out that it should not be difficult to integrate our continual prediction framework in an EHR system because the trained logistic regression model is only a mathematical equation (this is also true for many other machine learning methods) which in order to make predictions only requires the latest values of patient variables and these values are already present in the EHR. The system can then be made to trigger an alarm whenever the AKI prediction probability exceeds the threshold set corresponding to the desired level of sensitivity and specificity.

### 4. Conclusions

We introduced a new framework of continual prediction from EHR data and applied it for predicting AKI. Instead of applying the trained model at a particular time as was done in the past, the continual prediction model was applied continually over entire hospital stay whenever any patient variable changed.

Our experiments on a large dataset showed that the model out-performed one-time prediction models. Unlike one-time prediction models, the continual prediction model can take into account latest values of variables as they dynamically change over hospital stay. The continual prediction model also circumvents the shortcomings faced by one-time prediction models of either being applied too early to accurately predict AKI or being applied too late to prevent AKI.

**Conflict of Interest**
None.

# Appendix

To see the difference between continual and one-time prediction frameworks, we have shown below the information gain statistic for all the features. First it is shown for the data that is used to train the one-time-at-24-hour prediction model and next for the data that is used to train the continual prediction model (information gain is a property of the data and is independent of any model). The Weka machine learning software was used to obtain these. A higher value of the information gain statistic for a feature means that the feature is more helpful in distinguishing between AKI and non-AKI examples. The features are shown sorted in the decreasing order of information gain. It can be observed that for the one-time prediction, the most important features are pre-admission comorbidities and pre-admission medications. The features that change over hospital stays are not important for a one-time prediction because such a model is supposed to make predictions far ahead in time (for example, AKI incidence may occur even after 120 hours from admission) and at 24 hours from admission these features are not helpful in distinguishing between AKI and non-AKI hospital stays (as indicated by their low information gain statistic). We obtained similar trend of information gain statistic for other one-time prediction models as well. This also shows why a later one-time prediction model is not necessarily better at predicting AKI that develop later (see paper's Table 4 last column which changes very little). On the other hand, for continual prediction the features that change during hospital stays are the most important features because they are helpful in distinguishing between AKI and non-AKI hospital stays.

1. Features given to the one-time prediction model that predicts at 24 hours from admission ranked by the information gain statistic. Preadmission features are the top features.

| Information Gain | Feature |
| --- | --- |
| 0.006935 | Preadmission Diuretics |
| 0.006867 | Preadmission Heart Failure |
| 0.006867 | Preadmission Congestive Heart Failure |
| 0.005041 | Prior AKI |
| 0.004813 | Preadmission Respiratory Failure |
| 0.003634 | Preadmission K Sparing |
| 0.002644 | Preadmission Radiocontrast Dyes |
| 0.002591 | Preadmission Diabetes |
| 0.002451 | Preadmission Sepsis |
| 0.002281 | Preadmission ACE Inhibitors or NSAIDS or Diuretics |
| 0.002082 | Preadmission ARB or ACE Inhibitors or NSAIDS or Diuretics |
| 0.001865 | Preadmission ACE Inhibitors |
| 0.001715 | Preadmission Coronary Artery Disease |
| 0.001107 | Preadmission Lipid Lowering Drugs |
| 0.000888 | BP Systolic |
| 0.000826 | Preadmission Hypertension |
| 0.000821 | Preadmission ARB |
| 0.000737 | Race |
| 0.000735 | Preadmission Thrombocytopenia |
| 0.000704 | BP Diastolic |
| 0.00041 | Alcohol |
| 0.000345 | Tobacco |
| 0.000327 | Gender |
| 0.000326 | Heart Rate |
| 0.000313 | BMI |
| 0.000281 | Preadmission Aminoglycosides |
| 0.00025 | BUN |
| 0.000223 | Hypertension |
| 0.000153 | Drug |
| 0.000147 | Preadmission Hypercalcemia |
| 0.000123 | Preadmission Acyclovir |
| 8.24E-05 | Preadmission Cisplatin |
| 7.97E-05 | Preadmission Disorders of Lipoid Metabolism |
| 5.26E-05 | Heart Failure |
| 5.26E-05 | Congestive Heart Failure |
| 4.46E-05 | AST |
| 4.11E-05 | Preadmission Rhabdomyolysis |
| 3.79E-05 | NSAIDS |
| 3.41E-05 | Diuretics |
| 2.63E-05 | Respiratory Failure |
| 1.89E-05 | Thrombocytopenia |
| 1.85E-05 | Sepsis |
| 1.7E-05 | Preadmission NSAIDS |
| 7.72E-06 | Troponin |
| 7.6E-06 | Pancreatitis |
| 6.18E-06 | Disorders of Lipoid Metabolism |
| 5.07E-06 | Hypercalcemia |
| 4.03E-06 | Preadmission Hyperlipidemia |
| 2.53E-06 | Rhabdomyolysis |
| 2.53E-06 | ACE Inhibitors |
| 1.19E-06 | Diabetes |
| 1.08E-06 | ARB or ACE Inhibitors or NSAIDS or Diuretics |
| 1.08E-06 | ACE Inhibitors or NSAIDS or Diuretics |
| 9.58E-07 | Radiocontrast Dyes |
| 2.63E-09 | Hyperlipidemia |
| 0 | Blood Bilirubin |
| 0 | Acyclovir |
| 0 | Platelets |
| 0 | Preadmission Pancreatitis |
| 0 | Coronary Artery Disease |
| 0 | Temperature |
| 0 | Cisplatin |
| 0 | Aminoglycosides |
| 0 | K Sparing |
| 0 | Lipid Lowering Drugs |
| 0 | ARB |
| 0 | Age |

2. Features given to the continual prediction model ranked by the information gain statistic. Features that change during hospital stays are the top features.

| Information Gain | Feature |
|---|---|
| 0.015545 | Diuretics |
| 0.014529 | BP Systolic |
| 0.012971 | BP Diastolic |
| 0.008671 | ACE Inhibitors or NSAIDS or Diuretics |
| 0.008189 | Heart Rate |
| 0.00816 | Preadmission Diuretics |
| 0.00748 | ARB or ACE Inhibitors or NSAIDS or Diuretics |
| 0.007039 | Preadmission Heart Failure |
| 0.007039 | Preadmission Congestive Heart Failure |
| 0.006287 | Congestive Heart Failure |
| 0.006287 | Heart Failure |
| 0.00597 | Temperature |
| 0.005515 | Prior AKI |
| 0.004994 | Preadmission Respiratory Failure |
| 0.004883 | BUN |
| 0.003912 | Preadmission K Sparing |
| 0.003288 | Platelets |
| 0.00309 | Preadmission Diabetes |
| 0.002884 | Preadmission Radiocontrast Dyes |
| 0.002803 | Preadmission ACE Inhibitors or NSAIDS or Diuretics |
| 0.002744 | Preadmission Sepsis |
| 0.002601 | Preadmission ARB or ACE Inhibitors or NSAIDS or Diuretics |
| 0.00222 | Preadmission ACE Inhibitors |
| 0.00206 | K Sparing |
| 0.001759 | Preadmission Coronary Artery Disease |
| 0.001608 | AST |
| 0.001371 | Preadmission Lipid Lowering Drugs |
| 0.001154 | Respiratory Failure |
| 0.001153 | Preadmission Hypertension |
| 0.001094 | BMI |
| 0.000995 | Preadmission ARB |
| 0.000986 | Sepsis |
| 0.000881 | Preadmission Thrombocytopenia |
| 0.000803 | Race |
| 0.000802 | Cisplatin |
| 0.000802 | ACE Inhibitors |
| 0.000416 | Acyclovir |
| 0.000403 | Alcohol |
| 0.000394 | Thrombocytopenia |
| 0.000377 | Tobacco |
| 0.000356 | Hyperlipidemia |
| 0.000339 | Radiocontrast Dyes |
| 0.000327 | Diabetes |
| 0.000295 | Blood Bilirubin |
| 0.000255 | Gender |
| 0.000244 | Preadmission Aminoglycosides |
| 0.000225 | Aminoglycosides |
| 0.000188 | Hypercalcemia |
| 0.000185 | Drug |
| 0.000139 | ARB |
| 0.000132 | NSAIDS |
| 0.00013 | Preadmission Acyclovir |
| 0.000128 | Preadmission Hypercalcemia |
| 0.000116 | Troponin |
| 7.82E-05 | Preadmission Disorders of Lipoid Metabolism |
| 6.82E-05 | Preadmission Rhabdomyolysis |
| 4.65E-05 | Preadmission Cisplatin |
| 2.88E-05 | Disorders of Lipoid Metabolism |
| 2.01E-05 | Rhabdomyolysis |
| 1.25E-05 | Hypertension |
| 7.27E-06 | Lipid Lowering Drugs |
| 6.95E-06 | Preadmission NSAIDS |
| 2.86E-06 | Coronary Artery Disease |
| 1.86E-06 | Preadmission Hyperlipidemia |
| 2.95E-07 | Pancreatitis |
| 0 | Preadmission Pancreatitis |
| 0 | Age |